\documentclass[aps,prl,groupedaddress,]{revtex4}
\usepackage{latexsym}
\usepackage[dvips]{graphicx}

\newcommand{\eq}{\begin{equation}}
\newcommand{\feq}{\end{equation}}
\newcommand{\eqn}{\begin{eqnarray}}
\newcommand{\feqn}{\end{eqnarray}}
\newcommand{\arr}{\begin{eqnarray*}}
\newcommand{\farr}{\end{eqnarray*}}
\newcommand{\beq}{\begin{equation}}
\newcommand{\eeq}{\end{equation}}
\newcommand{\bea}{\begin{eqnarray}}
\newcommand{\eea}{\end{eqnarray}}

\def\beq{\begin{equation}}
\def\eeq{\end{equation}}
\def\feq{\end{equation}}
\def\bea{\begin{eqnarray}}
\def\eea{\end{eqnarray}}
\def\bc{\begin{displaymath}}
\def\ec{\end{displaymath}}
\def\lb{\label}

\def\la{\lambda}

\def\La{\Lambda}

\def\lb{\label}

\begin{document}


\title{Conformal symmetry of gravity and the cosmological
constant problem}

\author{Mariano Cadoni}
\email{mariano.cadoni@ca.infn.it}
\affiliation{Dipartimento di Fisica,
Universit\`a di Cagliari, and INFN sezione di Cagliari, Cittadella
Universitaria 09042 Monserrato, ITALY}


\begin{abstract}
 In
absence of matter Einstein gravity with a cosmological constant $\La$ can
be formulated as a scale-free theory depending only on the
dimensionless coupling constant $G \La$, where $G$ is Newton constant.
We derive the conformal field theory (CFT)
and its improved stress-energy tensor that describe the 
dynamics of conformally flat perturbations of the metric. The CFT has 
the form of a constrained $\lambda \phi^{4}$ field theory.
In the  cosmological framework the model describes the
usual Friedmann-Robertson-Walker flat universe. The conformal
symmetry of the gravity sector is broken by  coupling with matter.
The dimensional coupling constants $G$ and $\Lambda$ are introduced
by different terms in this coupling. If the vacuum of quantum matter fields
respects the symmetry of the gravity sector, the vacuum energy has
to be zero and the ``physical'' cosmological constant is generated
by the coupling of gravity with matter. This could explain the tiny
value  of the observed energy density driving the accelerating
expansion of the universe.

\end{abstract}

\maketitle
 The cosmological constant problem (see for reviews
\cite{Weinberg:1988cp,Carroll:2000fy,Peebles:2002gy,
Padmanabhan:2002ji,Ellwanger:2002cd,Nobbenhuis:2004wn}) arises in the
interplay between particle physics and cosmology. The problem is
originated by the attempt to explain the accelerating expansion of
the universe, which emerges from the recent remarkable observational
discoveries
\cite{Riess:1998cb,Perlmutter:1998np,Netterfield:2001yq,Knop:2003iy,
Tonry:2003zg,Spergel:2003cb,Tegmark:2003uf, Riess:2004nr}, in terms
of fundamental laws of microphysics. The observational data can be
explained introducing a cosmological constant $\Lambda$ ( a constant
curvature term) in the Einstein equations corresponding to a energy
density $\rho^{(obs)}\sim (10^{-3} eV)^{4}$ (Unless explicitly stated,
throughout this paper we will use natural units). On the other
hand, if we identify  this energy density with the vacuum energy of
ordinary particle physics and believe general relativity (GR) to
hold up to Planck scale, we expect $\rho^{(guess)}\sim
10^{120}\rho^{(obs)}$. We are therefore faced with the following
problems: $a)$ Why is the observed energy density so tiny with
respect to that expected from microphysics? $b)$ Why  is  $\rho$ not
exactly equal to zero?\footnote{ This two questions are referred to
as the old cosmological constant problem. A third puzzling point
with which we are faced when trying to understand the universe in
which we live is the coincidence problem: Why is the vacuum energy
density today  of the same order of magnitude as that of matter? In
this paper we will not address this question.}

During the last two decades  many  proposal have been put forward to
solve the cosmological constant problem (see e.g. Ref.
\cite{Weinberg:1988cp,Carroll:2000fy,Peebles:2002gy,
Padmanabhan:2002ji,Ellwanger:2002cd,Nobbenhuis:2004wn}): symmetry
principles, environmental selection, back-reaction, breaking down of
general relativity, large extra dimensions, dynamical dark energy.
Despite these efforts, none of the proposal can be considered
satisfactory. It seems by now evident that the origin of the
cosmological constant problem is a very subtle conspiracy between physics at
short and large scales. If the energy density $\rho^{(obs)}$  is
originated by  a local quantum field theory,  it has to be
associated to a length scale
$L_{s}\sim(\hbar/\rho^{(obs)}c)^{1/4}¥\sim 10^{-2} cm$ (note that we
have reinstated units). This in turn implies the emergence of new
physics already at  length scales  $L_{s}$, which is very unlikely,
although not completely ruled out by the experiments. Any
modification of the Einstein action - either by short-distance terms
(e.g coming
 from string theory) or by large extra dimensions - are strongly
constrained because they are not allowed to modify the gravitational interaction
in the range millimeters-astronomical distances.
The tiny value of $\rho^{(obs)}$
could be also  traced back to an infrared modification of general relativity
at length
scales $L_{S}\sim (c^{2}/G\rho^{(obs)})^{1/2}¥\sim  10^{29}cm$.
But in this case, it is very difficult to understand how the
physics at those scales could prevent the vacuum energy density of quantum
fields to  become comparable with $\rho^{(guess)}$.

In view of this very intricate situation a proper understanding of
the role played by the cosmological constant in the Einstein theory
of gravity  may be very useful and indicate promising directions for
solving the problem. The key idea, which will be the
starting point of this paper, is to question the standard
geometrical interpretation of $\Lambda$ and to interpret it from a
``field theoretical'' point of view as a {\sl coupling constant}.
This approach has been first proposed by Polyakov \cite{Polyakov:2000fk} and more 
recently in Ref. \cite{Jackiw:2005yc}. It has been noted that the conformal mode 
of Einstein gravity can be described by a $\lambda \phi^{4}$ field 
theory \cite{Polyakov:2000fk,Jackiw:2005yc}. It has been also proposed that the quantum triviality 
of the $\lambda \phi^{4}$ theory  could be responsible for infrared 
screening, driving to zero the value of the cosmological constant 
\cite{Polyakov:2000fk,Jackiw:2005yc}.

In this paper we will elaborate further  on this idea.
We will show that in the absence of matter Einstein
gravity with a cosmological constant is a scale-free theory. 
We derive the conformal field theory (CFT) in four dimensions (4D) 
and its improved stress-energy tensor that describe the 
dynamics of conformally flat perturbations of the metric. The CFT has 
the form of a constrained $\lambda \phi^{4}$ field theory.
In the  cosmological framework the theory describes the
usual Friedmann-Robertson-Walker flat universe. 
 The conformal
symmetry of the gravity sector is broken by  coupling with
matter. The coupling constants $G$ and $\Lambda$ are introduced by
different terms in this coupling. If the vacuum of matter fields
respects the symmetry of the gravity sector, the vacuum energy has
to be zero and the ``physical'' cosmological constant is generated
by the coupling of gravity with matter.

In the Einstein theory of gravity, described by the action
\beq\lb{action}
S={1\over 16\pi G}\int d^{4}x \sqrt{-g}\left( R - 2\Lambda\right),
\eeq
$\Lambda$ appears as a free parameter, which is usually interpreted,
geometrically, as a constant curvature term of the spacetime
independent from the presence of matter sources.
This geometrical interpretation  is, from the physical point of view,
rather unsatisfactory.
First, if the spacetime has a curvature in absence of matter-energy
sources, what is the physical origin of the curvature?
Second,
in the cosmological framework the term looses its geometrical interpretation
and becomes a (rather exotic) form of matter, with constant density
and equation of state $p=-\rho$.
Third,  the cosmological constant problem amounts to explain why the
effective cosmological constant $\Lambda_{eff}= \Lambda - 8\pi
G\langle \rho_{M}¥\rangle$ is so close to zero.
But the cancellation required here is rather unnatural.
It requires to sum a  free
parameter with something determined by the matter sources to produce
a small number.

The above difficulties do not arise if we forget  the geometrical
meaning of $\Lambda$ and interpret $\Lambda$ and $G$ on equal
footing, as  {\sl coupling constants}. The second step is to use
standard dimensional analysis (see for instance Ref.
\cite{deAlfaro:1979hg} for the case of gravity with
zero cosmological constant)  to realize that the gravitational field
$g_{\mu\nu}$ is not canonically defined. In fact its group
dimensionality,  which characterizes the behavior under dilatations,
is $-2$, whereas its engineering dimensionality is $0$. A canonical
defined metric tensor is obtained by  the redefinition
\beq\lb{red}
g_{\mu\nu}\to \frac{g_{\mu\nu}}{\la^{2}},
\feq
where $\La=3\la^{2}$.
With the field redefinition (\ref{red})  the Einstein action
(\ref{action}) becomes scale-free and depends only on the
dimensionless coupling $\alpha_{0}= G\la^{2}$
\beq\lb{action1}
S={1\over 16\pi \alpha_{0}}\int d^{4}x \sqrt{-g}\left( R - 6\right).
\eeq
For field theories in flat space-time independence of the action
from dimensional constants is related to scale and, more in general,
conformal invariance. For  diffeomorphisms invariant theories such as
general relativity, this relationship becomes more involved
\cite{deAlfaro:1979hg}. In the case of  Einstein gravity with a cosmological
constant this scale symmetry can be easily identified as a subgroup
of the isometry group of the de-Sitter spacetime, which is  solution
of the theory. We note in passing, that  the dS/CFT (AdS/CFT)
correspondence \cite{Maldacena:1997re,Strominger:2001pn} is also
based on the conformal isometry of the de Sitter (anti de-Sitter)
spacetime.

The scale symmetry is hidden in the action (\ref{action}) because of
the presence of the dimensional coupling constants $G,\la$, but
becomes explicit in the action (\ref{action1}). Scale or more in
general, conformal, symmetry has been already used in the past for
trying to solve the cosmological constant problem
\cite{Zee:1981ff,Booth:2002hg,Barbour:2002ad,Mannheim:1999bu}.
Usually one modifies the Einstein action, either by considering
scalar-tensor theories or quadratic curvature terms, in order to
construct a scale invariant action. Unfortunately, these
modifications affect also the behavior of general relativity
 on distance scales where it  works   perfectly. In this paper we
 will not try to modify GR but we will select the regimes of
 the theory when conformal invariance manifests itself explicitly.

The investigation of the scale symmetry of the action (\ref{action1})
is rather complicated because of general covariance and the presence
of many degrees of freedom. In particular,  it is well known that
scale invariance is not compatible with the general covariance of
general relativity. The Einstein-Hilbert term $\sqrt{-g} R$ in the
action (\ref{action}) is not scale invariant. In order to circumvent
this problem we  consider as in Ref. \cite{Polyakov:2000fk,Jackiw:2005yc} only 
conformally flat perturbation of the metric, i.e 
we fix the diffeomorphisms in the action and
consider only one gravitational degree of freedom, the conformal
factor of the metric:
\beq\lb{cf}
g_{\mu\nu}=\psi^{2}\eta_{\mu\nu}.
\feq 
This is a rather drastic simplification, which we expect to
apply only to configurations  with a high degree of symmetry. But it
allows to formulate a theory of gravity as a field theory in flat
spacetime and  makes evident the conformal properties of Einstein
gravity coupled with a cosmological constant.  Clearly,  we pay a
high  price,  we loose completely the general covariance of the
theory and we throw away most  of modes of the gravitational field.
As a consequence we are forgetting the metric nature of the field
$\psi$ and we are able to describe only a very limited part of
gravitational physics. For instance, we cannot describe  simple
gravitational structures such as a black hole or global features of
the spacetime such as cosmological horizons. On the other hand there
are at least two regimes in which the gravitational field admits the
description given by Eq. (\ref{cf}). The first is cosmology,
 a spatially flat Robertson-Walker universe. The second is the local
 behavior of the gravitational field. Locally, we can always turn
 off the gravitational field and set $g_{\mu\nu}=\eta_{\mu\nu}$.
 Obviously flat spacetime is not a exact solution of Einstein gravity with
 a cosmological constant. However, we can consider it as an
 approximate solution that holds when we consider gravitational
 physics at  length scales $<<L_{\La}=\La^{-1/2}$.
 In this case the term $\La g_{\mu\nu}$ in the Einstein equations can
 be neglected and flat spacetime becomes  a solution of the theory.
 In terms of the ansatz (\ref{cf}) this behavior corresponds to an 
 almost constant field $\psi$.
 It is also interesting to notice that our proposal is complementary
 to the so-called  unimodular theories of gravity \cite{Unruh:1988in}.  In these
 theories the  conformal factor of the metric is held fixed
 and the cosmological constant appears as an integration
 constant of the gravitational field equations.

Using Eq. (\ref{cf}) into the action (\ref{action1}), we get
\beq\lb{cfa} S={3\over 8\pi \alpha_{0}}\int d^{4}x \left[
(\partial\psi)^{2} - \psi^{4}¥\right]. \eeq
This is a CFT in flat space-time, which has the form of a $\lambda 
\phi^{4}$ field theory.
Because we are fixing the
diffeomorphisms of the action (\ref{action1}) we expect the appearance
of constraints. In fact one can show, after some manipulation, that the
Einstein equations stemming from the action (\ref{action1}) and
evaluated with the ansatz (\ref{cf})  are
equivalent to the field equations derived from (\ref{cfa}) constrained
by the equation
\beq\lb{set}
T^{(\psi)}_{\mu\nu}= \psi\left(\partial_{\mu}\partial_{\nu}\psi
-\frac{1}{4}\eta_{\mu\nu}¥ \partial^{2}\psi\right)
-2\left(\partial_{\mu}\psi\partial_{\nu}\psi
-\frac{1}{4}\eta_{\mu\nu} (\partial\psi)^{2}¥\right)=0
\feq
The tensor $T^{(\psi)}_{\mu\nu}$ can be interpreted as the improved
stress-energy tensor of the 4D CFT defined by the action (\ref{cfa}).
In fact $T^{(\psi)}_{\mu\nu}$ is identically traceless and it is
conserved, i.e satisfies $\partial_{\mu}T^{(\psi)\mu\nu}=0$, by
virtue of the field equations
\beq\lb{fe}
\partial^{2}\psi+2\psi^{3}=0,
\feq
coming from the action (\ref{cfa}).
We have therefore shown that the dynamics of the conformal factor of the
metric in Einstein gravity with a cosmological constant can be
described by  a 4D CFT constrained by a vanishing stress-energy tensor.
The action (\ref{cfa}) is invariant under the action of the 4D
conformal group generated by the infinitesimal transformations
\beq\lb{it}
\xi^{\mu}= a^{\mu}+\omega^{\mu}_{\nu}x^{\nu}-\epsilon
x^{\mu}+c^{\mu}x^{2}-2 x^{\mu} c_{\nu}¥x^{\nu},
\feq
where $a^{\nu}, \omega^{\mu}_{\nu}, \epsilon, c^{\nu}$ generate
respectively,  translations, rotations, dilatations and special
conformal transformations.  The field $\psi$ transforms as a
conformal field with weight $-1$:
\beq\lb{tf}
\delta\psi=-\left( \xi^{\nu}\partial_{\nu}+\frac{1}{4} (\partial 
\xi)\right)\psi.
\feq

It is well know that the conformal symmetry puts stringent
constraints on the form of the correlations functions of the theory.
In particular,  the solutions of the field equations (\ref{fe}) are
$\psi=1/\sqrt{(2x^{2})}$ (the so-called meron type solutions
\cite{deAlfaro:1978dz,deAlfaro:1979hg}). Eq. (\ref{set}) puts
additional constraints on the form of the solution, so that only the
solution $\psi= 1/t$ is allowed. Meron type solutions are an
example of classical scale-free solutions.
In the context of Einstein gravity with vanishing cosmological
constant they have been used some time ago to argue about the
short-distance independence of gravity from Newton constant
\cite{deAlfaro:1979hg}.

The conformal symmetry of the gravitational sector  is in general
broken by coupling with matter fields $\chi^{M}$ and we expect that
the dimensional coupling constants $G,\la$ are introduced by this
breaking. The gravity-matter action becomes  $S_{tot}=
S+S_{M}¥(\chi^{M}, g_{\mu\nu})$, where $S$ is the gravitational
action (\ref{action}) and $S_{M}$ is the action for matter fields.
Using Eq. (\ref{cf}) the action becomes 
\beq\lb{act2}
S_{tot}={3\over 8\pi \alpha_{0}}\int d^{4}x \left[
(\partial\psi)^{2} - \psi^{4}\right]+ S_{M}(\psi,\chi^{M},\la), 
\eeq
whereas the constraint (\ref{set}) is now 
\beq\lb{con1}
T^{(\psi)}_{\mu\nu}= -4\pi
G\left(T^{M}_{\mu\nu}-\frac{1}{4}\eta_{\mu\nu}T^{M}¥\right)\psi^{2}¥,
\feq 
where $T^{M}_{\mu\nu}=T^{M}_{\mu\nu}(\psi,\chi^{M},\la)$ is the
usual matter stress-energy tensor (evaluated for the  metric
(\ref{cf})) and $T$ is its (flat spacetime) trace. The pattern of
the conformal symmetry breaking   depends on the form of the matter
action $S_{M}(\psi,\chi^{M},\la)$. In general we will have $S_{M}=
S_{0}(\chi^{M})+ \la^{-2}\psi^{2}S_{1}(\chi^{M})+ \ldots
\la^{-2n}\psi^{2n}S_{n}(\chi^{M})+\ldots.$ and  corresponding terms
in the stress-energy tensor $T^{M}_{\mu\nu}$.

The term $S_{0}$ corresponds to a conformally invariant matter
 action, which has $T^{M}=0$. In this case the field equations for $\psi$
\beq\lb{fep}
\partial^{2}\psi+2\psi^{3}=\frac{4}{3} \pi G T^{M} \psi,
\feq remain conformal invariant and it is the constraint
(\ref{con1}) which breaks the symmetry and introduces in the theory
the dimensional coupling $\alpha_{1}= \alpha_{0}\la^{-2}=G$ (unless
the right hand side of Eq.(\ref{con1}) is identically zero). The
symmetry breaking term $S_{1}$ introduces the coupling constant
$\alpha_{1}$  in the field equations (\ref{fep}) and in the
constraint (\ref{con1}). The term $S_{2}$ requires the introduction
of the coupling constant $\alpha_{2}=
\alpha_{0}\la^{-4}=G\la^{-2}¥$. Generically, the term $S_{n}$
introduces in the theory the coupling constant $\alpha_{n}=
\alpha_{0}\la^{-2n}= G\la^{-(2n-2)}$. Notice that the coupling
constants $\alpha_{0}, \alpha_{1}¥\ldots\alpha_{n}¥$ define a
hierarchy of coupling constants of dimensionality $[L]^{0}, [L]^{2},
\ldots [L]^{2n}$. The terms introduced in the action by $\alpha_{n}$
dominate at large scales with respect to those introduced by
$\alpha_{m}$ ($m<n)$. Conversely, the terms introduced by
$\alpha_{m}$ dominate at short scales. We therefore have 
scale-free, short-distance, gravitational physics characterized
by $\alpha_0$. The intermediate (from terrestrial to astronomical
distances) behavior  breaks the conformal symmetry and is determined
by $\alpha_1$. The long-distance (cosmological scales) behavior
breaks also conformal symmetry and is characterized by the coupling
constant $\alpha_2$.

We can easily apply  our approach to cosmology. In fact in the case
of a spatially flat Robertson-Walker universe  the metric  can be put
in the form (\ref{cf}), just by introducing the conformal time $t$. The
field $\psi$ describes the radius of the universe and is a function
of $t$ only. For matter in the form of a perfect fluid characterized
by pressure $p$ and density $\rho$ equations (\ref{fep}) and (\ref{con1}) 
 become respectively
\bea\lb{ceq}
\ddot \psi -2\psi^{3}&=& - \frac{4}{3}
\pi\alpha_{2}(3p-\rho)\psi^{3}¥\nonumber\\
\ddot \psi- 2\frac{(\dot \psi)^{2}}{\psi}&=& - 4
\pi\alpha_{2}(p+\rho) \psi^{3}. \eea 
The previous equation are
equivalent to the usual Friedmann equations, we just need to change
the time variable  $dt=dt'/\psi$ and to rescale $\psi$. Notice that
the consistency condition between Equations (\ref{ceq}) maintains the 
form  it has for the Friedmann equations:  $\dot \rho=-3
(p+\rho)(\dot\psi/\psi)$.

Written in the  form (\ref{ceq}), the  cosmological equations
have  a lot of
interesting physical features which are completely obscured  in the usual
formulation.
First of all,  equations (\ref{ceq}) depend on the coupling constant
$\alpha_{2}$ but not on $\alpha_{1}=G$. This means that we are
considering gravitation at (large) length scales where $G$ becomes
irrelevant and the dynamics is determine completely by $G/\la^{2}$.
At smaller length-scales (e.g solar system distances) gravitational
physics is determined by $\alpha_{1}=G$ and $\alpha_{2}$ becomes
irrelevant.
Moreover, $\la$ appear in the equations  as a coupling constant, in the
same way as $G$.
This solves the dichotomy mentioned at the beginning of this paper
concerning the interpretation of $\La$ both as a geometrical term
independent of the distribution of matter and as a exotic form of
matter. In our formulation the  energy density driving the
accelerating  expansion of the universe can be considered just as a
form of matter with equation of state $p_{a}¥=-\rho_{a}¥$. The cosmological
constant $\la$ builds up with $G$ the strength $\alpha_{2}$  of its gravitational
interaction.

 The cosmological equations (\ref{ceq}) also have a very
nice physical interpretation. The first equation describes the
dynamics of the conformal mode of the metric. This dynamics is
conformally symmetric either when matter is absent ($p=\rho=0$) or
when the matter dynamics is also conformal ($p=\rho/3$). The second
equation in (\ref{ceq}), or equivalently Eq. (\ref{con1}),
gives a relationship between the
stress-energy of  the conformal mode $\psi$ and the stress-energy
tensor of matter. This equal footing description of sources and
gravitational field is rather unusual in gravity theories. Using the
consistency condition the same equation can be written as a general
relation between the energy density of the conformal mode $\psi$ and
the flux of matter: 
\beq
\lb{e1} T^{(\psi)}_{00}= -3 \pi\alpha_{2}¥
(p+\rho)\psi^{4}= \pi \alpha_{2} \frac{\psi^{5}}{\dot\psi}
\dot\rho.
\feq
For matter with equation of state $ p=-\rho$ both the
flux of matter and the energy density of the conformal mode are
zero. For $p>-\rho$ ($p<-\rho$)  both the flux of matter and the
energy density of $\psi$ are negative (positive).

Let us now discuss the implication of our  results for the
cosmological constant problem. We have argued above that the
gravitational field allows for a parametrization  as in Eq.
(\ref{cf})  in two regimes: its behavior on length scales $<<
L_{\La}$ and cosmology. These are two limiting regimes, but are the
relevant ones for the cosmological constant problem. In fact  this
problem arises from the interplay between micro- and macro-physics.
We do not expect that gravitational physics at intermediate scales -
such as astronomical distances - could have some impact on the
problem. If our description applies at short distances, the
conformal symmetry could be responsible for  driving to zero the  
vacuum energy of quantum fields. A mechanism to achieve this goal has 
been proposed in Ref. \cite{Jackiw:2005yc}.  It is based on the quantum triviality 
of the $\lambda \phi^{4}$ theory  that could drive  to zero the 
value of the cosmological constant. 

The constant energy density driving the accelerating
expansion of the universe could be generated by  the coupling with
matter, which breaks the conformal invariance (see Eq. (\ref{con1}),
(\ref{fep})). More in detail this energy density is generated by the
terms in the matter action that depend on $\psi^{4}¥$ and couple to
gravity trough the coupling constant $\alpha_{2}$. Because, the
origin of this energy density is the coupling with ordinary matter
we naturally expect it to be of order $\rho^{(obs)}$.

The key question to be answered now is: how much is our approach
reliable? After all we are describing gravity in a very rough way
and we are spoiling it by most of the nice features of the metric
theories of gravity. Moreover, the mechanism proposed in 
Ref. \cite{Polyakov:2000fk,Jackiw:2005yc} that drives 
the cosmological constant to zero has more or less the status of a 
conjecture and should be made more precise.
On the other hand our approach makes 
the conformal invariance of Einstein gravity with a cosmological
constant explicit. Conformal invariance can be hardly reconciled  with
the general covariance of the theory. Some fixing of the
diffeomorphisms invariance of GR is therefore required in order to
make the scale symmetry explicit. 
It is obvious that these
investigations have to be considered as a starting point for trying
to build a theory of pure gravity  that is manifestly scale invariant
while maintaining most of the features of GR, at least at
astronomical length scales. Because the main results of our paper
are based on symmetry arguments we expect them to remain true for
such would-be theory. In view of the above-mentioned incompatibility
between scale symmetry and general covariance, the task of building
such a theory is far from being trivial. It is likely that deep
modifications of GR will be required.

Another point, which deserves interest is the physical origin of the coupling
constant $\alpha_{2}$ (or equivalently $\la$) and  of the
energy density $\rho_{a}¥$ with equation of state $p_{a}¥=-\rho_{a}¥$.
$\alpha_{2}$ is a free parameter, so that in order to determine both
$\alpha_{2}$ and $\rho_{a}¥$ we need beside the observed value
$\rho^{(obs)}$ other independent observational data involving them.
Of course  this requires   detailed information about
the origin of of $\rho_{a}¥$. We  just know that $\rho_{a}$ is
generated by terms in the matter action depending on $\psi^{4}$.
For instance such term can be generated by the potential of a scalar
field coupled to gravity. More in general they could be generated by
global effects of the gravitational field and/or of the quantum
vacuum of matter fields for which our simplified description does not
apply.

\end{document}